\documentclass{article}
\usepackage[english]{babel}

\usepackage[letterpaper,top=2cm,bottom=2cm,left=3cm,right=3cm,marginparwidth=1.75cm]{geometry}

\usepackage{csquotes}
\usepackage{amsmath}
\usepackage{graphicx}
\usepackage{authblk}
\usepackage{subcaption}
\usepackage{multirow}
\usepackage{booktabs}
\usepackage{array}
\usepackage{indentfirst}

\usepackage[colorlinks=true, allcolors=blue]{hyperref}
\title{Translating Multimodal AI into Real-World Inspection: TEMAI Evaluation Framework and Pathways for Implementation}
\author[1,3]{Zehan LI}
\author[1,2]{Jinzhi Deng}
\author[1,2]{Haibing Ma\thanks{To whom correspondence should be addressed; Email:mahaibing@moximize.ai}}
\author[1]{Chi Zhang}
\author[1]{Dan Xiao}
\affil[1]{Moximize.ai}
\affil[2]{Shanghai Zhongqiao Vocational And Technical University}
\affil[3]{China Creative Studies Institute}
\begin{document}
\maketitle
\begingroup
\renewcommand{\thefootnote}{}
\footnotetext{LZH@moximize.ai(ZeHan Li); dengjinzhi@moximize.ai(Jinzhi Deng); mahaibing@moximize.ai(Haibing Ma); zhangchi@moximize.ai(Chi Zhang); xiaodan@moximize.ai(Dan Xiao)}
\endgroup

\begin{abstract}
This paper introduces the Translational Evaluation of Multimodal AI for Inspection (TEMAI) framework, bridging multimodal AI capabilities with industrial inspection implementation. Adapting translational research principles from healthcare to industrial contexts, TEMAI establishes three core dimensions: Capability (technical feasibility), Adoption (organizational readiness), and Utility (value realization). The framework demonstrates that technical capability alone yields limited value without corresponding adoption mechanisms. TEMAI incorporates specialized metrics including the Value Density Coefficient and structured implementation pathways. Empirical validation through retail and photovoltaic inspection implementations revealed significant differences in value realization patterns despite similar capability reduction rates, confirming the framework's effectiveness across diverse industrial sectors while highlighting the importance of industry-specific adaptation strategies.
\end{abstract}

\providecommand{\keywords}[1]
{
  \small	
  \textbf{Keywords:} #1
}

\section{Introduction}
\setlength{\parindent}{0cm}
Industrial inspection tasks are fundamental to ensuring operational continuity and safety in manufacturing sectors, serving as a cornerstone for preventive maintenance and risk mitigation. These tasks, however, are plagued by systemic inefficiencies, including labor-intensive workflows, hazardous working environments (e.g., high-temperature zones or toxic gas exposure), and heavy reliance on empirical knowledge that is difficult to standardize or transfer across industries\cite{ref1}. Despite incremental advancements in automation technologies—such as drones, AR-assisted devices, and IoT-enabled sensors—the integration of these tools into inspection workflows has yielded limited returns due to fragmented deployment, high implementation costs, and insufficient interoperability between hardware and software systems \cite{ref2}. For instance, while drones have reduced human exposure to dangerous environments in power grid inspections, their operational scope remains constrained by battery life and data processing bottlenecks\cite{ref3}.

\setlength{\parindent}{2em}
Advances in LLMs and LLM-powered systems offer promising opportunities for transforming industrial inspection workflows. Research indicates that LLM-powered software could significantly reduce task completion time across various work activities, suggesting substantial potential for improving inspection efficiency\cite{ref4}. This is particularly relevant for inspection processes requiring data analysis, documentation, and knowledge transfer. Recent advancements in algorithmic robustness and context-aware reasoning, are now approaching the threshold for deployment in industrial production scenarios requiring high-precision computational tasks \cite{ref5, ref6}. A comparative study between leading AI models (GPT-4 and GLM-4) has demonstrated a 30 to 100-fold increase in operational efficiency when applying GenAI to complex computational tasks, suggesting similar potential gains in industrial inspection applications \cite{ref7}.

Large language models (LLMs), when integrated with advanced sensing networks that combine visual, thermal, acoustic, and other data streams, enable comprehensive inspection capabilities including real-time anomaly detection, predictive maintenance, and adaptive decision support—features critical for addressing the dynamic challenges of industrial inspection \cite{ref8}. Nevertheless, enterprise adoption of AI-powered inspection solutions remains cautious. High-stakes investment risks, compounded by historically underwhelming ROI from earlier automation initiatives, deter organizations from committing to large-scale AI deployments \cite{ref9}. This reluctance stems from a lack of standardized frameworks to quantify both tangible economic benefits (e.g., Cost Reduction, Efficiency Enhancement) and intangible value (e.g., ESG Value Creation, Risk Prevention), as well as the absence of clear implementation roadmaps to navigate technical and organizational hurdles. Therefore, a systematic method for translating multimodal AI into real-world inspection is required. 

\section{Methodology}

\subsection{Translational Research Foundation}

\setlength{\parindent}{0cm}
Originally proven successful in healthcare settings, Translational Research (TR) methodology establishes a systematic knowledge transfer framework that effectively addresses various challenges in the transition from laboratory development to practical deployment \cite{ref10,ref11}. This methodology has demonstrated unique advantages in converting multimodal AI into industrial inspection applications \cite{ref12}. 

\setlength{\parindent}{2em}
The transition from healthcare to industrial inspection contexts is particularly appropriate given several key parallels: both domains require high-precision systems with significant safety implications, face similar regulatory scrutiny, and involve complex decision-making processes based on multimodal data interpretation. Additionally, both fields share common implementation barriers including the need for specialized expertise, organizational reluctance to new technologies, and difficulty in quantifying return on investment for advanced AI systems.

Inspired by the TEHAI (Translational Evaluation of Healthcare AI) framework — a mature tool for evaluating clinical AI applications \cite{ref13, ref14},  we propose the TEMAI (Translational Evaluation of Multimodal AI for Inspection) framework, the first cross-industry value assessment system tailored for AI-powered inspection.  To achieve this objective, we established a three-level hierarchical framework. The highest level comprises core dimensions, representing fundamental evaluation aspects; the middle level consists of components, detailing specific elements within each dimension; and the lowest level contains assessment criteria, providing measurable indicators. It is an evaluation framework that maintains theoretical integrity while ensuring practical applicability.

\subsection{Framework Development Process}

\setlength{\parindent}{0cm}
The framework development followed a systematic three-phase approach aligned with our core evaluation dimensions. In Phase I, we conducted comprehensive domain analysis to establish theoretical foundations, particularly focusing on translatable elements from healthcare AI evaluation frameworks to industrial inspection contexts.

\setlength{\parindent}{2em}
Phase II involved expert panel consultation and framework refinement. A panel of ten domain experts was carefully selected with a balanced composition of 4:3:3 ratio (AI/ML specialists : industrial inspection experts : ESG evaluation professionals). This ratio was determined based on the relative complexity and technical depth of the three core dimensions - Capability, Adoption, and Utility. AI/ML specialists primarily validated the Capability dimension, industrial inspection experts focused on Adoption considerations, while ESG professionals contributed to the Utility assessment framework.

The expert consensus process employed the Analytic Hierarchy Process (AHP) methodology through three rounds of Delphi studies:

(1) Round 1: Initial framework structure validation

(2) Round 2: Component-level assessment criteria refinement

(3) Round 3: Weight determination for each dimension and its sub-components, including the development of critical multiplicative indices such as the Value Density Coefficient (VDC), Technical Absorption Capacity index, and other specialized metrics for multimodal AI evaluation

The final framework underwent iterative refinement based on pilot implementation feedback from two manufacturing facilities, ensuring its practical applicability while maintaining theoretical rigor. This systematic approach enabled the development of a comprehensive evaluation framework that effectively addresses the technical, operational, and strategic aspects of multimodal AI implementation in industrial inspection scenarios.

\subsection{Value Quantification Framework}

\setlength{\parindent}{0cm}
The framework incorporates a standardized value quantification approach through a man-hour calculation model: $Man-hour =  BaseRate \times AIEfficiency \times RiskWeight$.

\setlength{\parindent}{2em}
The $BaseRate$ represents industry-specific labor cost benchmarks, while $AIEfficiency$ comprises weighted scores from the core dimensions ($a+b+c=1$). The $RiskWeight$ factor accounts for implementation complexity and potential risks. Each core dimension's score derives from its corresponding components and assessment criteria, ensuring comprehensive evaluation while maintaining hierarchical integrity.

This standardized value quantification framework provides the critical theoretical foundation for TEMAI implementation pathways (Section 4). By translating complex multimodal AI system performance into measurable human labor equivalents, the framework enables organizations to conduct accurate return-on-investment calculations and achieve real-time economic value accounting , thereby facilitating data-driven implementation decisions.

\subsection{Specialized Metric Development Methodology}

\setlength{\parindent}{0cm}
The specialized indices throughout the TEMAI framework were systematically developed using an extended Delphi approach with structured expert elicitation \cite{ref15}, necessitated by the absence of established benchmarks at the intersection of multimodal AI and industrial inspection.

\setlength{\parindent}{2em}
For metric development, we assembled an expanded panel of 18 experts: six AI/ML specialists (multimodal systems), five industrial inspection professionals, four ESG measurement experts, and three technology adoption specialists. Each metric underwent a four-stage process: (1) Conceptual Formulation establishing theoretical grounding; (2) Component Identification determining constituent variables; (3) Quantification Protocol development; and (4) Validation and Calibration with sensitivity analysis.

The implementation incorporated methodological rigor through: quantitative inter-expert agreement assessment using Kendall's W coefficient (consensus threshold $W\geq 0.7$) \cite{ref16}, stability analysis between Delphi rounds, cross-validation between complementary indices, and boundary condition identification through sensitivity analysis. This approach aligns with established practices in technology assessment research \cite{ref17} and has precedent in framework development for complex socio-technical systems where purely empirical approaches are premature \cite{ref15}.

For complex indices like the Value Density Coefficient, the panel employed structured decomposition to identify core components (task criticality, knowledge concentration, risk exposure) before establishing measurement protocols through consensus-building. Each component metric in subsequent sections was derived through this methodology, ensuring both theoretical soundness and practical applicability in real-world industrial inspection contexts.

\section{TEMAI Framework: A Three-Dimensional Evaluation System for Industrial AI Inspection}

\subsection{Framework Overview}
\begin{figure}
\centering
\includegraphics[width=1\linewidth]{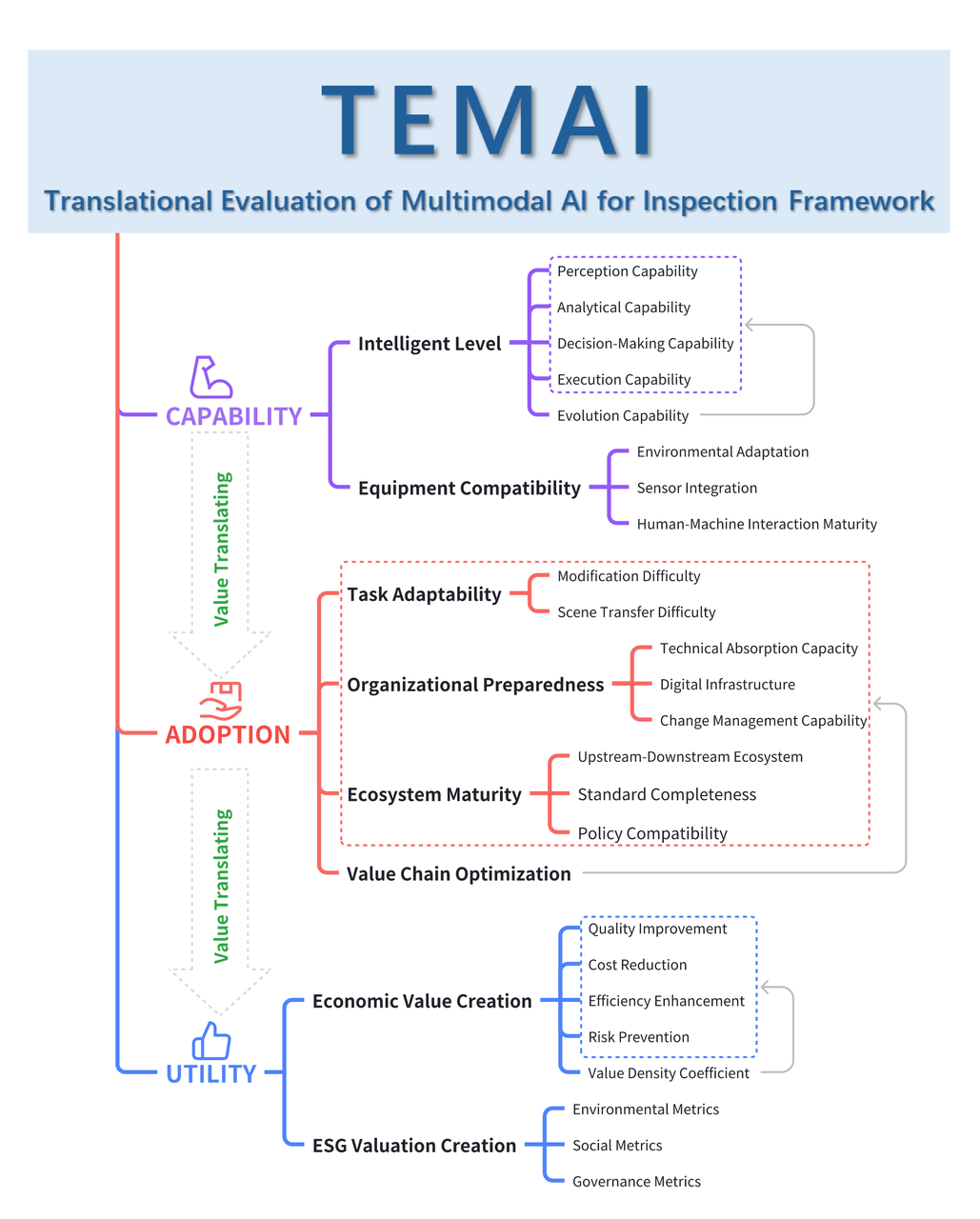}
\caption{\label{fig:TEMAI framework}TEMAI framework—Capability, Utility, and Adoption.}
\end{figure}

\setlength{\parindent}{0cm}
Outlined in Figure \ref{fig:TEMAI framework} , there are three Core Dimensions(Level 1) in TEMAI framework—Capability, Utility, and Adoption. They are further articulated through 8 Components (Level 2) and 24 Assessment Criteria (Level 3). Components and their associated Assessment Criteria belonging to the same Core Dimension are represented in the same color for visual distinction. Assessment Criteria with cross-Component relationships are linked by arrows, illustrating multiplicative effects within the framework. 

\subsection{Core Dimensions and Their Components}
\subsubsection{Capability Dimension}
\setlength{\parindent}{0cm}
The Capability dimension evaluates the operational adaptability and technical performance of multimodal AI systems in industrial inspection scenarios. This dimension is particularly relevant for multimodal AI systems that must interpret complex, heterogeneous data streams from various sensors simultaneously. The objective is to assess not merely individual modality performance but the system's integrated capabilities across the full inspection workflow.

\setlength{\parindent}{2em}
Within the Capability dimension, we identify two key components. The first component, Intelligent Level, encompasses four assessment criteria that form a progressive capability chain. Perception Capability measures the accuracy and speed with which the multimodal system detects defects or anomalies across different data streams, such as identifying surface imperfections through visual channels while simultaneously detecting thermal anomalies. Analytical Capability assesses how effectively the system interprets and correlates data across modalities, for example, determining whether a visual defect corresponds with an abnormal thermal signature. Decision-Making Capability evaluates the system's ability to prioritize findings and recommend appropriate responses based on multi-modal analytical outcomes. Execution Capability focuses on how effectively the system translates decisions into actionable outcomes in industrial environments, including integration with robotic systems or human-in-the-loop workflows.

A critical assessment criterion within the Capability dimension is the Evolution Capability, which represents the system's ability to improve over time through continuous learning and adaptation. Evolution Capability supports Perception through improved scene adaptation across different inspection environments, enhances Analysis via cross-domain reasoning that leverages patterns identified across multiple modalities, and optimizes Decision-making and Execution through dynamic response optimization. This adaptive capability is essential for industrial inspection applications where conditions frequently change and inspection requirements evolve over time.

The second component, Equipment Compatibility, addresses the physical and operational integration of multimodal AI systems within existing industrial infrastructure. This includes Environmental Adaptation, which evaluates the system's resilience across diverse operational conditions typical in industrial settings, such as varying lighting, temperature fluctuations, or high-noise environments. Sensor Integration assessment ensures seamless interaction between AI systems and various sensing technologies ranging from specialized cameras to thermal imagers and acoustic sensors. The evaluation focuses on how effectively multiple sensor types can be integrated to provide comprehensive inspection data while maintaining data quality and synchronization. Human-Machine Interaction Maturity assesses the quality and effectiveness of interfaces between AI inspection systems and human operators.

\subsubsection{Adoption Dimension}
\setlength{\parindent}{0cm}
The Adoption dimension evaluates how effectively multimodal AI inspection systems are integrated into industrial workflows by assessing organizational readiness, task adaptability, and ecosystem maturity. This dimension addresses the human, structural, and systemic factors that either facilitate or impede successful implementation of advanced inspection technologies.

\setlength{\parindent}{2em}
The first component, Task Adaptability, measures how well multimodal AI solutions can be customized for specific industrial inspection requirements. This includes evaluating Modification Difficulty—the effort required to adapt the system to existing workflows—and Scene Transfer Difficulty, which assesses the system's portability across different operational environments. In multimodal contexts, this adaptability extends to how effectively the system incorporates and prioritizes different data streams based on task-specific requirements.

The second component, Organizational Preparedness, evaluates whether human resources, infrastructure, and operational processes are sufficiently developed to support multimodal AI implementation. This includes assessing Technical Absorption Capacity—the organization's ability to understand and effectively utilize complex multimodal data outputs. Digital Infrastructure evaluates the readiness of underlying technical systems to support AI operations. Change Management Capability assesses the organization's ability to navigate the transformational aspects of adopting advanced inspection technologies.

The third component, Ecosystem Maturity, examines the development of supporting networks essential for sustainable AI implementation. This includes the Upstream-Downstream Ecosystem, which evaluates the readiness of suppliers and customers to integrate with AI-enhanced inspection processes. Standards Completeness assesses the maturity and availability of industry standards that support interoperability and best practices for multimodal systems. Policy Compatibility measures the alignment between AI solutions and regulatory frameworks.

The fourth component, Value Chain Optimization, serves as a critical element within the Adoption dimension. This component focuses on transforming traditional industrial processes into AI-enhanced workflows that maximize value creation. Value Chain Optimization contributes to Task Adaptation through streamlined integration processes, enhances Organizational Preparedness through optimized resource allocation, and supports Ecosystem Maturity by creating incentives for partner development.

\subsubsection{Utility Dimension}
\setlength{\parindent}{0cm}
The Utility dimension measures the tangible and intangible value created by multimodal AI inspection systems, focusing on economic outcomes and contributions to broader ESG (Environmental, Social, and Governance) goals. This dimension provides a comprehensive assessment framework for quantifying return on investment and societal impact.

\setlength{\parindent}{2em}
The first component, Economic Value Creation, encompasses four primary assessment criteria that holistically evaluate financial impact. Quality Improvement measures how multimodal inspection enhances product quality through more comprehensive defect detection than single-modality systems can provide. Cost Reduction quantifies operational savings from automated inspection processes that combine multiple sensor types to reduce manual inspection requirements. Efficiency Enhancement evaluates productivity gains from accelerated inspection speeds and reduced false positives through cross-modal validation. Risk Prevention assesses the system's contribution to avoiding costly failures, recalls, or safety incidents through early detection of complex defect patterns across multiple data streams.

A distinctive feature of the Utility dimension is the Value Density Coefficient, which provides a measurement of value concentration within inspection applications. This coefficient represents the concentrated value creation potential within specific inspection applications, reflecting how the same multimodal AI capability might generate substantially different value depending on critical factors such as application criticality, failure cost magnitude, or regulatory significance. The Value Density Coefficient relates to quality, cost, efficiency, and risk metrics while also integrating ESG considerations into holistic value assessment.

The second component, ESG Value Creation, evaluates broader impacts beyond immediate economic returns. Environmental Metrics assess how improved inspection efficiency contributes to waste reduction, energy conservation, and sustainable resource utilization. Social Metrics evaluate safety improvements, working condition enhancements, and skill development opportunities created through AI implementation. Governance Metrics examine transparency improvements, compliance assurance, and risk management capabilities enabled by comprehensive multimodal inspection data.

\section{TEMAI Implementation Pathways: From Theory to Practice}
\setlength{\parindent}{0cm}
While the TEMAI framework provides comprehensive assessment dimensions, implementation requires practical tools and processes. Different industry sectors face varying regulatory intensity and government support levels, which influence implementation priorities and resource allocation. Section 5.3 provides a detailed quadrant analysis of these external factors to help adapt implementation approaches to specific industry environments.

\subsection{Case Study: Multimodal AI Implementation in Retail Store Inspection}
\setlength{\parindent}{0cm}
Retail store inspection represents a distinct implementation domain for multimodal AI technology with unique operational constraints and value creation opportunities. This case study examines the application of our TEMAI framework to evaluate multimodal AI implementation in retail environments, where inspection tasks typically include merchandise display compliance, pricing accuracy, inventory management, and health and safety adherence.

\subsubsection{Implementation Context and Technical Configuration}
\setlength{\parindent}{0cm}
The retail inspection case study involved deploying a multimodal AI system integrating visual, textual, and spatial data streams to monitor store conditions across multiple locations. The system architecture incorporated:
\setlength{\parindent}{2em}
High-resolution cameras capturing visual data for planogram compliance and product presentation

OCR-capable sensors for price tag validation and promotional material verification

Spatial mapping sensors establishing store layout conformity with corporate standards

Backend multimodal LLM processing capabilities for integrated analysis and reporting

\begin{figure}[htbp]
    \centering
    \begin{subfigure}[b]{0.3\textwidth}
        \centering
        \includegraphics[width=\textwidth]{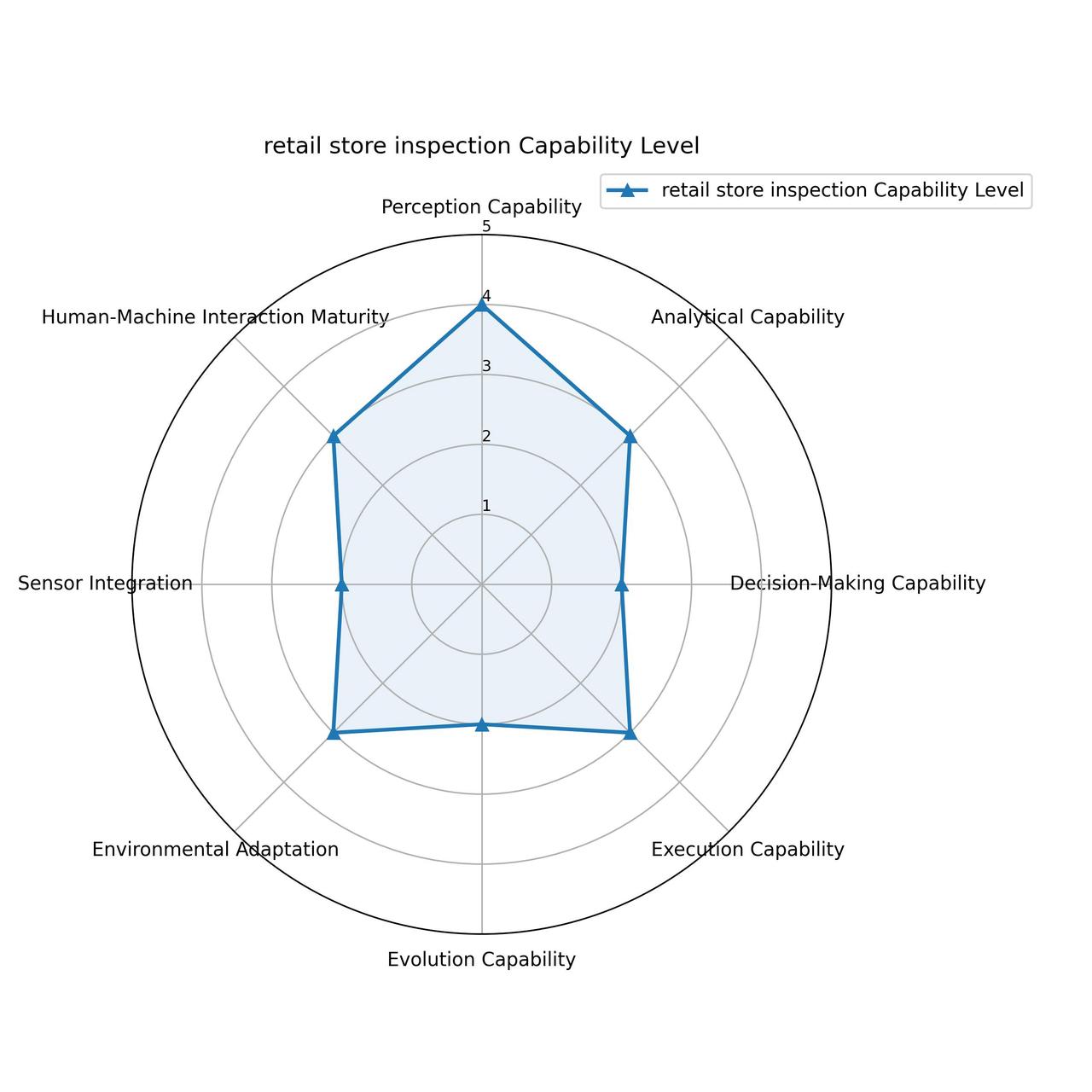}
        \caption{}
        \label{fig:retail_c_l}
    \end{subfigure}
    \hfill
    \begin{subfigure}[b]{0.3\textwidth}
        \centering
        \includegraphics[width=\textwidth]{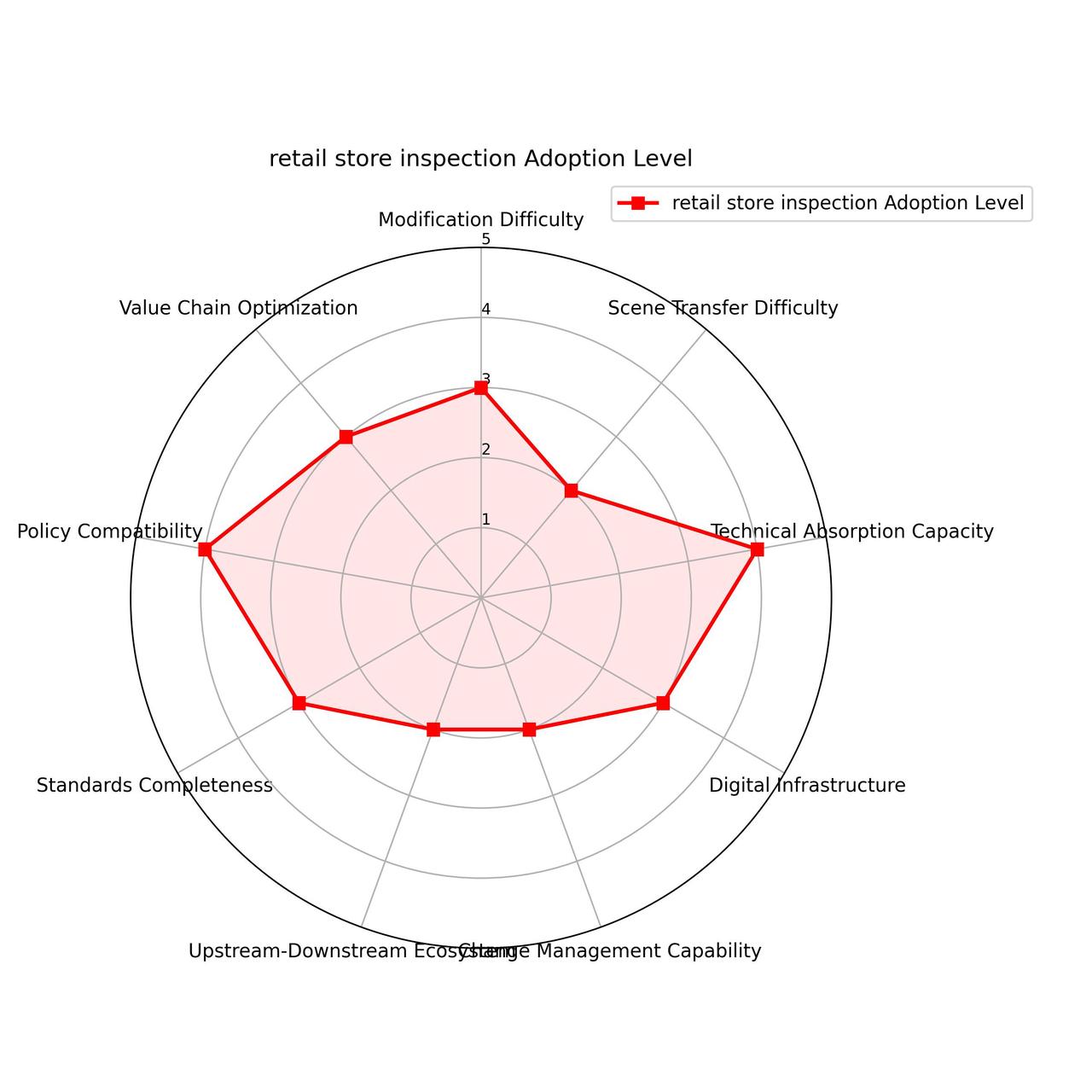}
        \caption{}
        \label{fig:retail_a_l}
    \end{subfigure}
    \hfill
    \begin{subfigure}[b]{0.3\textwidth}
        \centering
        \includegraphics[width=\textwidth]{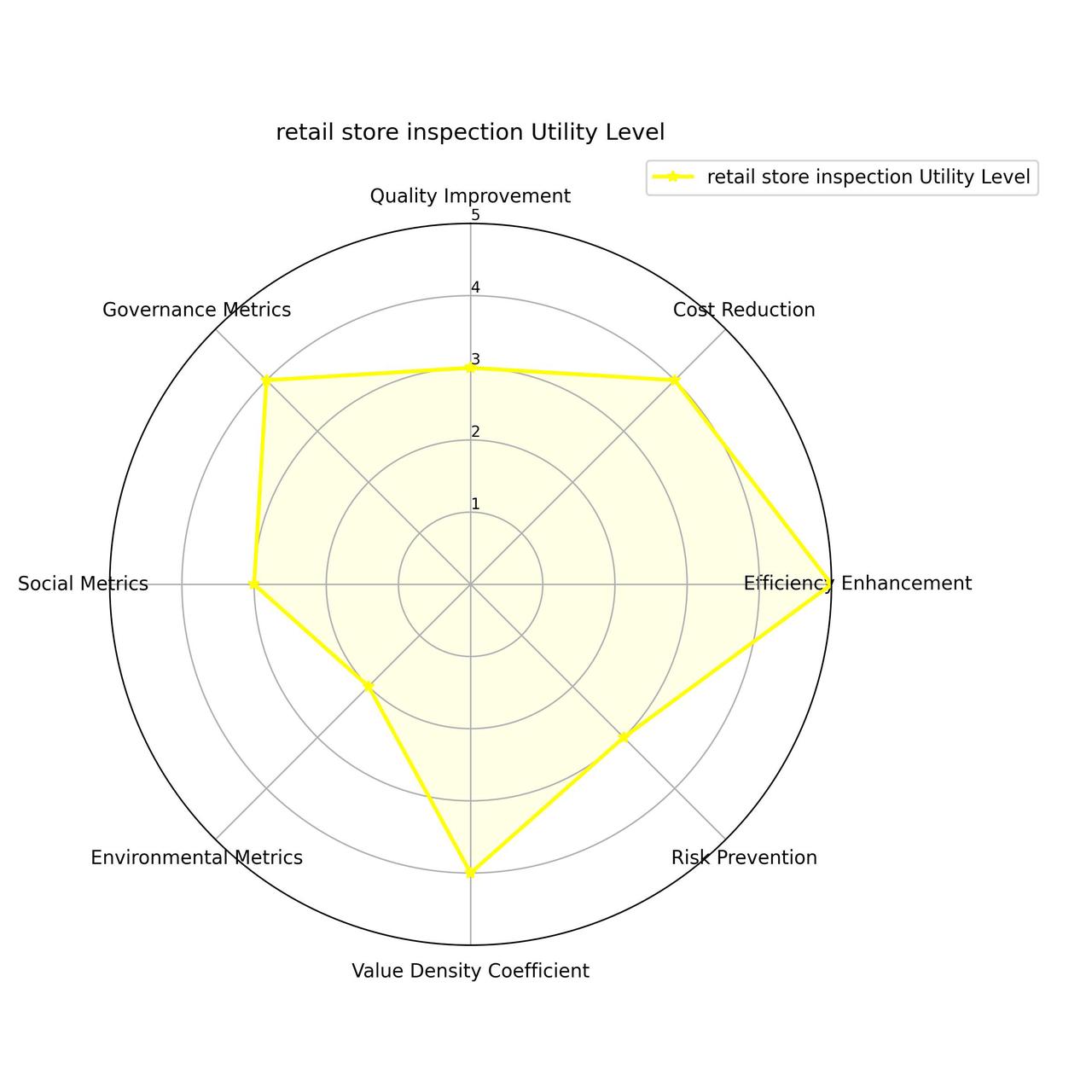}
        \caption{}
        \label{fig:retail_u_l}
    \end{subfigure}
    \caption{Pre-conversion Scores—Capability, Utility, and Adoption in retail inspection}
    \label{fig:retail_level}
\end{figure}

Figure \ref{fig:retail_level} presents the expert panel's evaluation of multimodal AI application in retail inspection across the three TEMAI dimensions. Using a five-level rating scale (1-5),Figure\ref{fig:retail_c_l} shows Capability assessment across 8 criteria, Figure\ref{fig:retail_a_l}  displays Adoption evaluation across 9 criteria, and Figure\ref{fig:retail_u_l}  illustrates Utility assessment across 8 criteria. Higher ratings indicate stronger performance in each dimension.

Expert panel assessment revealed moderate capability scores across most dimensions, with Perception Capability receiving the highest rating (Level 4, score 80) due to the system's effective visual analysis in controlled indoor environments. In the TEMAI framework, pre-conversion scoring maps Level 1 to 20 points, Level 2 to 40 points, Level 3 to 60 points, Level 4 to 80 points, and Level 5 to 100 points. Decision-Making Capability received notably lower ratings (Level 2, score 40), reflecting limited autonomous decision authority in retail contexts where human judgment remains preferred for customer-facing remediation actions.

\subsubsection{TEMAI Evaluation Results for Retail Implementation}
\setlength{\parindent}{0cm}
The TEMAI framework assessment yielded an initial Capability score of 57.56, reflecting the multimodal system's moderate technical readiness for retail inspection tasks. The Adoption conversion rate was calculated at 51.16 \%, resulting in an effective Capability score of 29.44 after implementation constraints. The Utility value conversion rate reached 70.46 \%, producing a final value realization score of 10.61.

\begin{figure}[htbp]
    \centering
    \begin{subfigure}[b]{0.3\textwidth}
        \centering
        \includegraphics[width=\textwidth]{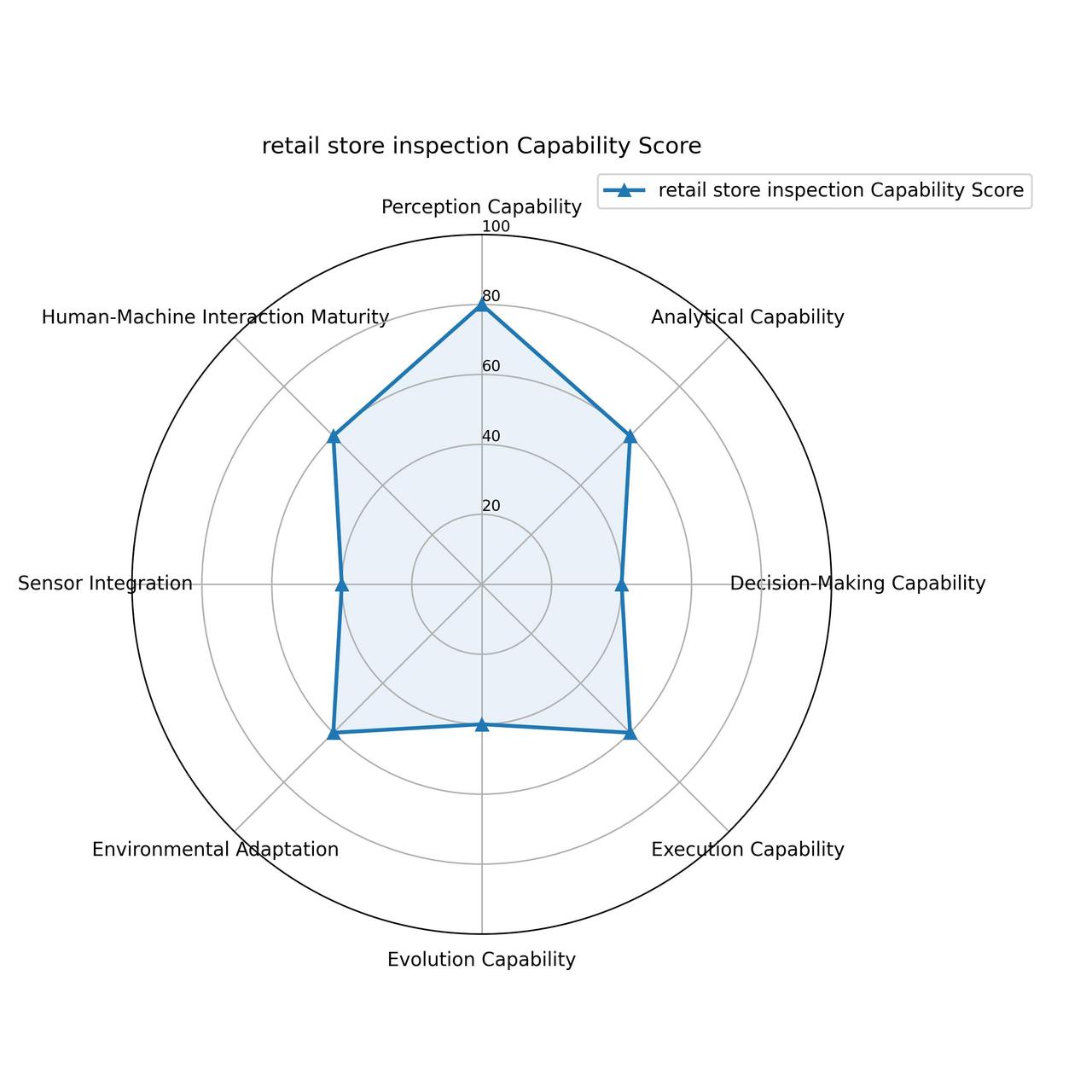}
        \caption{}
        \label{fig:retail_c_s}
    \end{subfigure}
    \hfill
    \begin{subfigure}[b]{0.3\textwidth}
        \centering
        \includegraphics[width=\textwidth]{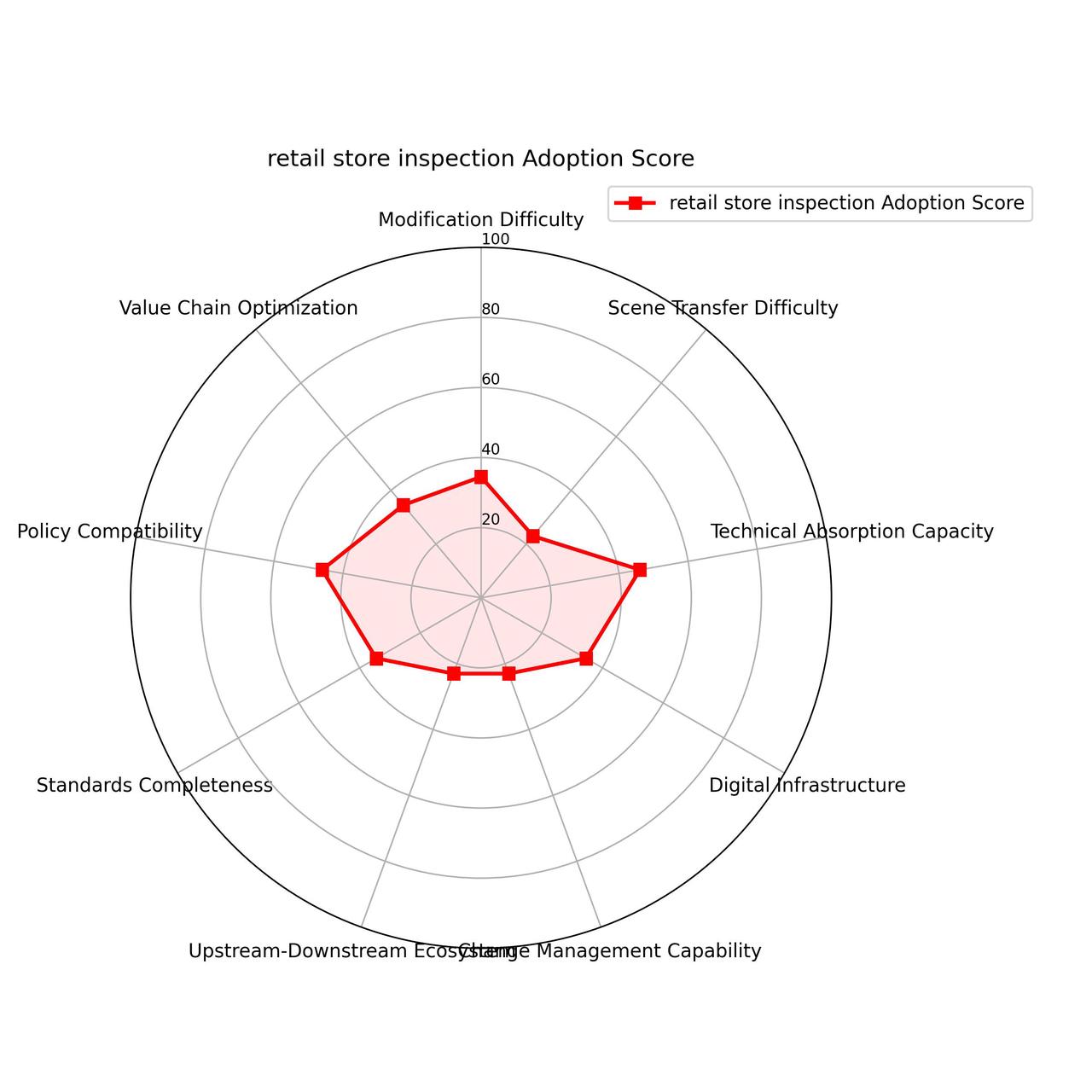}
        \caption{}
        \label{fig:retail_a_s}
    \end{subfigure}
    \hfill
    \begin{subfigure}[b]{0.3\textwidth}
        \centering
        \includegraphics[width=\textwidth]{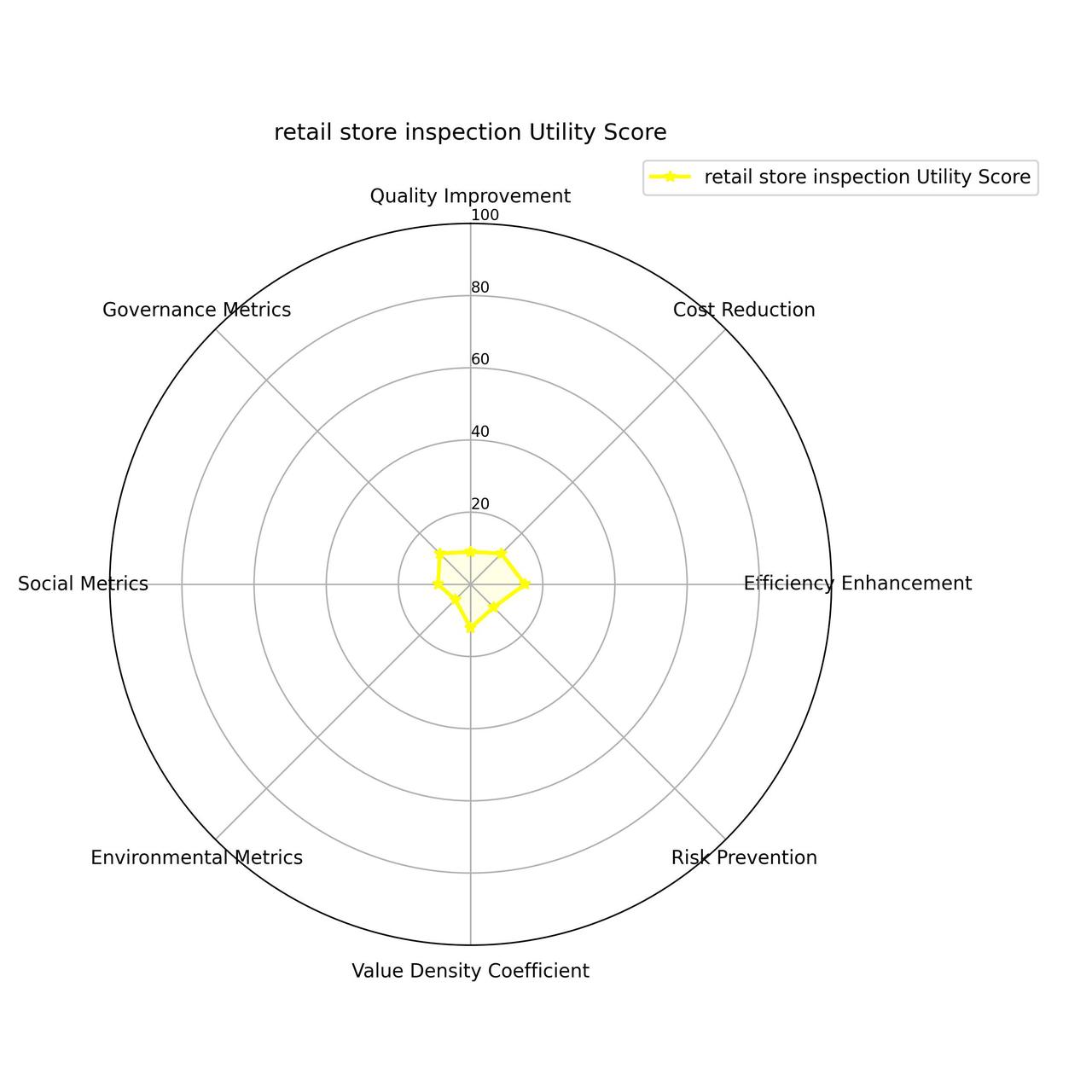}
        \caption{}
        \label{fig:retail_u_s}
    \end{subfigure}
    \caption{Converted Scores—Capability, Utility, and Adoption in retail inspection}
    \label{fig:retail_score}
\end{figure}

\setlength{\parindent}{2em}
Figure \ref{fig:retail_score} illustrates scores (0-100 scale) across implementation stages in retail inspection. Figure\ref{fig:retail_c_s} shows the original AI capability score and composition, Figure\ref{fig:retail_a_s} presents transformed capabilities during deployment with conversion factors, and Figure\ref{fig:retail_u_s} displays the final value realization score and its components.

Analysis of the component-level results revealed several critical implementation considerations:

In the Capability dimension, the lower scores for Evolution Capability (Level 2, score 40) and Sensor Integration (Level 2, score 40) highlighted adaptation challenges in retail environments where layout changes are frequent and sensor deployment faces aesthetic constraints. The Human-Machine Interaction Maturity score (Level 3, score 60) reflected adequate but not exceptional interface design for store personnel with varying technical expertise.

The Adoption dimension revealed significant implementation barriers, particularly in Scene Transfer Difficulty (Level 2, the converted score approximately 20, for the calculation formula, please refer to the appendix) due to high variability between store layouts and merchandise arrangements. Change Management Capability (Level 2,the converted score approximately 20) likewise indicated organizational resistance to technology-driven workflow changes among established retail staff. However, Technical Absorption Capacity (Level 4, the converted score approximately 50) demonstrated retail organizations' existing familiarity with data-driven operations.

The Utility dimension showed strongest performance in Efficiency Enhancement (Level 5, the converted score approximately 15), reflecting the system's ability to dramatically reduce manual inspection time. Cost Reduction received positive assessment (Level 4, the converted score approximately 10) due to labor savings, while Governance Metrics (Level 4, the converted score approximately 10) highlighted compliance benefits. Environmental Metrics received the lowest utility rating (Level 2, the converted score approximately 8), indicating limited perceived environmental value creation in this application context.

\subsubsection{Implementation Challenges and Adaptation Strategies}
\setlength{\parindent}{0cm}
The retail case implementation revealed several practical challenges requiring specific adaptation strategies:

\setlength{\parindent}{2em}
Store-to-Store Variability: High variance in store layouts necessitated extensive model retraining and configuration adjustments, addressed through implementation of self-adaptive reference models that could calibrate to varying store conditions.

Staff Technology Interaction: Variable technical proficiency among retail personnel required development of simplified interfaces and task-specific training modules to ensure consistent system utilization across locations.

Visual Obstructions: Customer traffic and merchandise relocations frequently created visual obstructions, requiring temporal data integration algorithms that could maintain inspection continuity despite intermittent sensor blockages.

Value Communication: Quantifying ROI proved challenging due to the preventive nature of many benefits. Implementation success required development of counterfactual analysis methods that could demonstrate avoided losses rather than solely direct savings.

\subsection{Case Study: Multimodal AI Implementation in Photovoltaic System Inspection}
\setlength{\parindent}{0cm}
Photovoltaic (PV) system inspection represents a substantially different implementation domain characterized by outdoor environmental conditions, critical safety considerations, and high-consequence failure modes. This case study examines the application of the TEMAI framework to evaluate multimodal AI implementation in PV inspection, where tasks include detecting module degradation, identifying hotspots, and assessing structural integrity.
\subsubsection{Implementation Context and Technical Configuration}

The PV inspection case study involved deploying a multimodal AI system integrating thermal imaging, visual inspection, electrical performance data, and geospatial information to comprehensively assess solar array conditions. The deployment architecture incorporated:
\setlength{\parindent}{2em}

Drone-mounted thermal cameras identifying temperature anomalies indicating potential hotspots;

High-resolution RGB cameras detecting visual defects including microcracks and discoloration;

I-V curve tracers capturing electrical performance metrics;

Weather data integration for environmental context;

Multimodal fusion algorithms correlating findings across data streams.

\begin{figure}[htbp]
    \centering
    \begin{subfigure}[b]{0.3\textwidth}
        \centering
        \includegraphics[width=\textwidth]{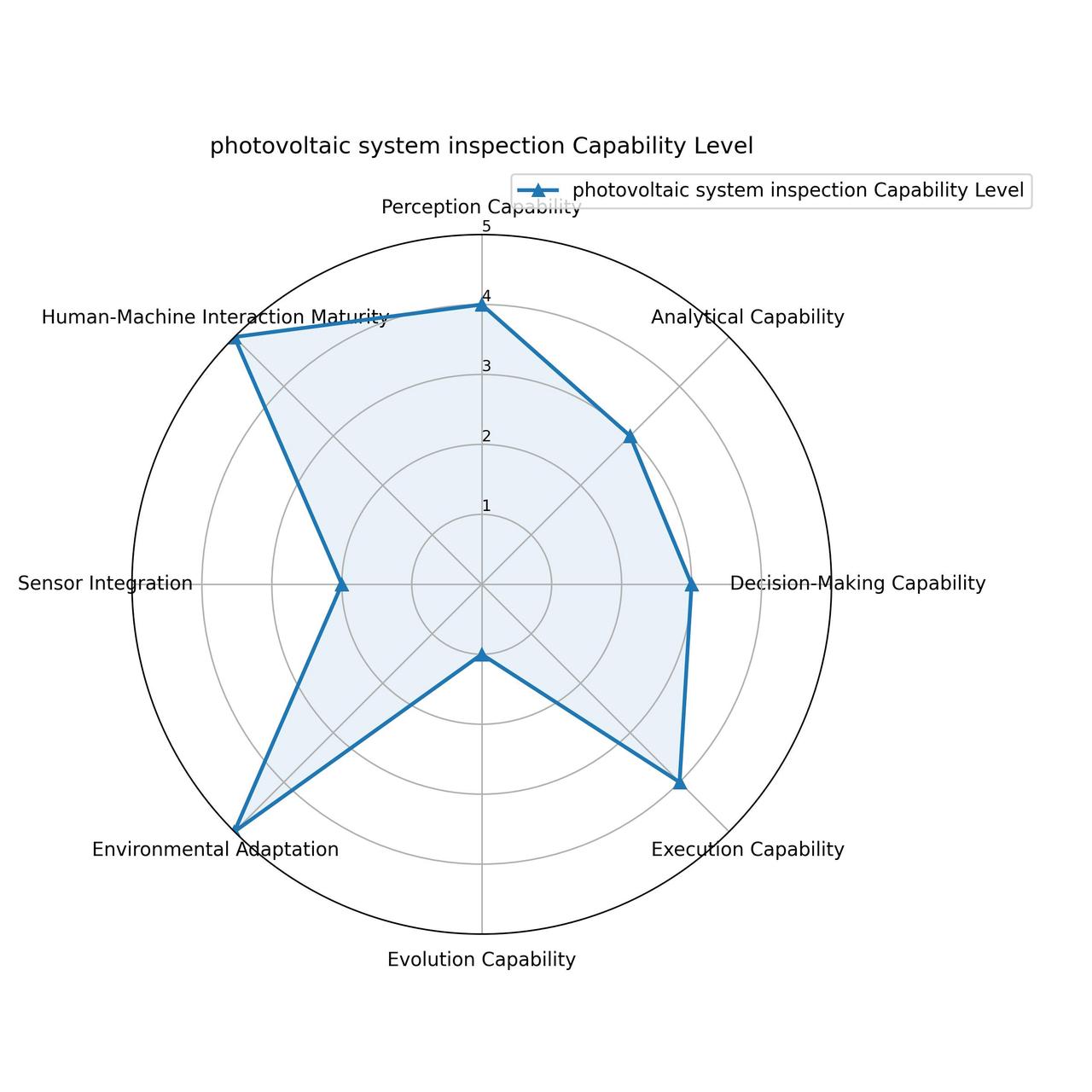}
        \caption{}
        \label{fig:pv_c_l}
    \end{subfigure}
    \hfill
    \begin{subfigure}[b]{0.3\textwidth}
        \centering
        \includegraphics[width=\textwidth]{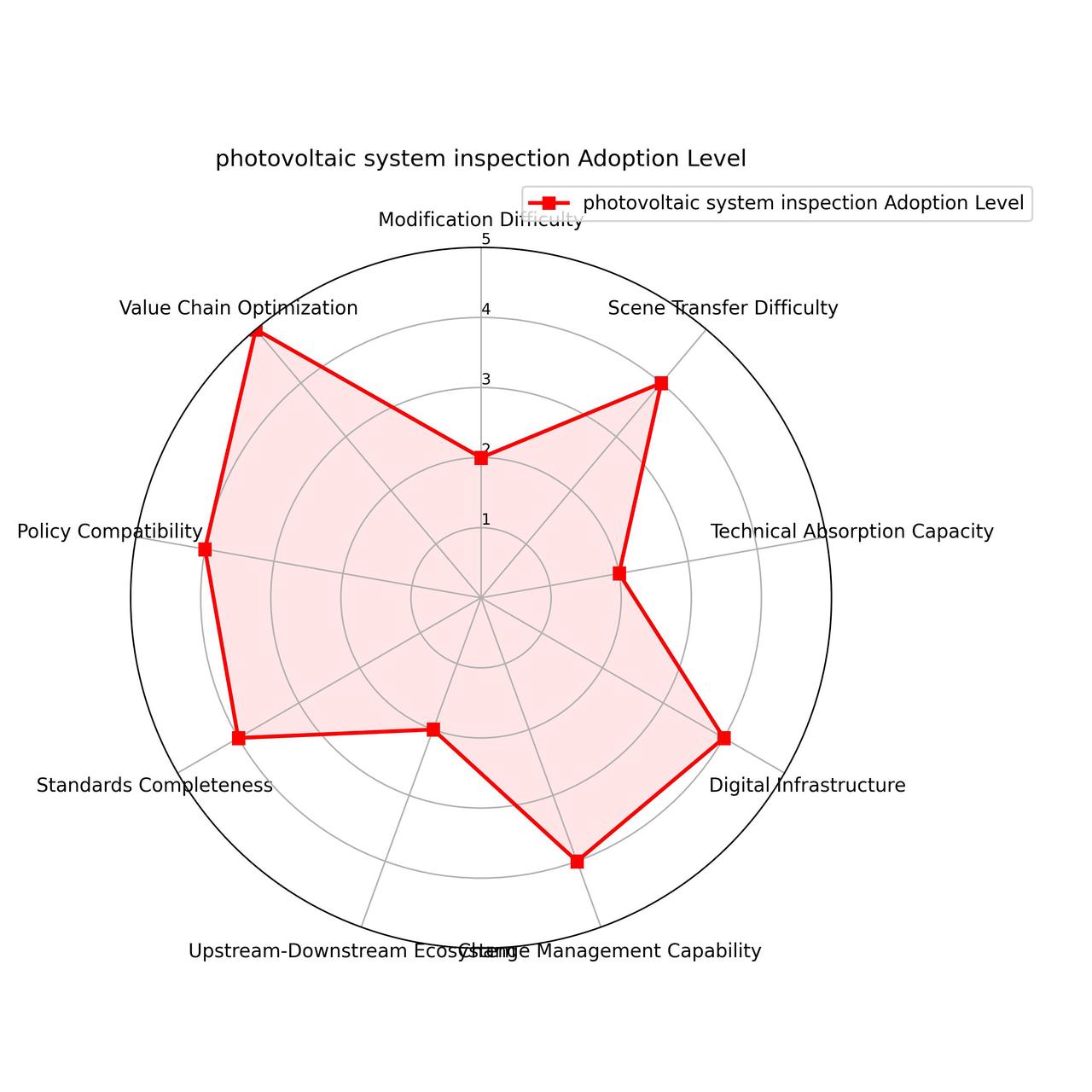}
        \caption{}
        \label{fig:pv_a_l}
    \end{subfigure}
    \hfill
    \begin{subfigure}[b]{0.3\textwidth}
        \centering
        \includegraphics[width=\textwidth]{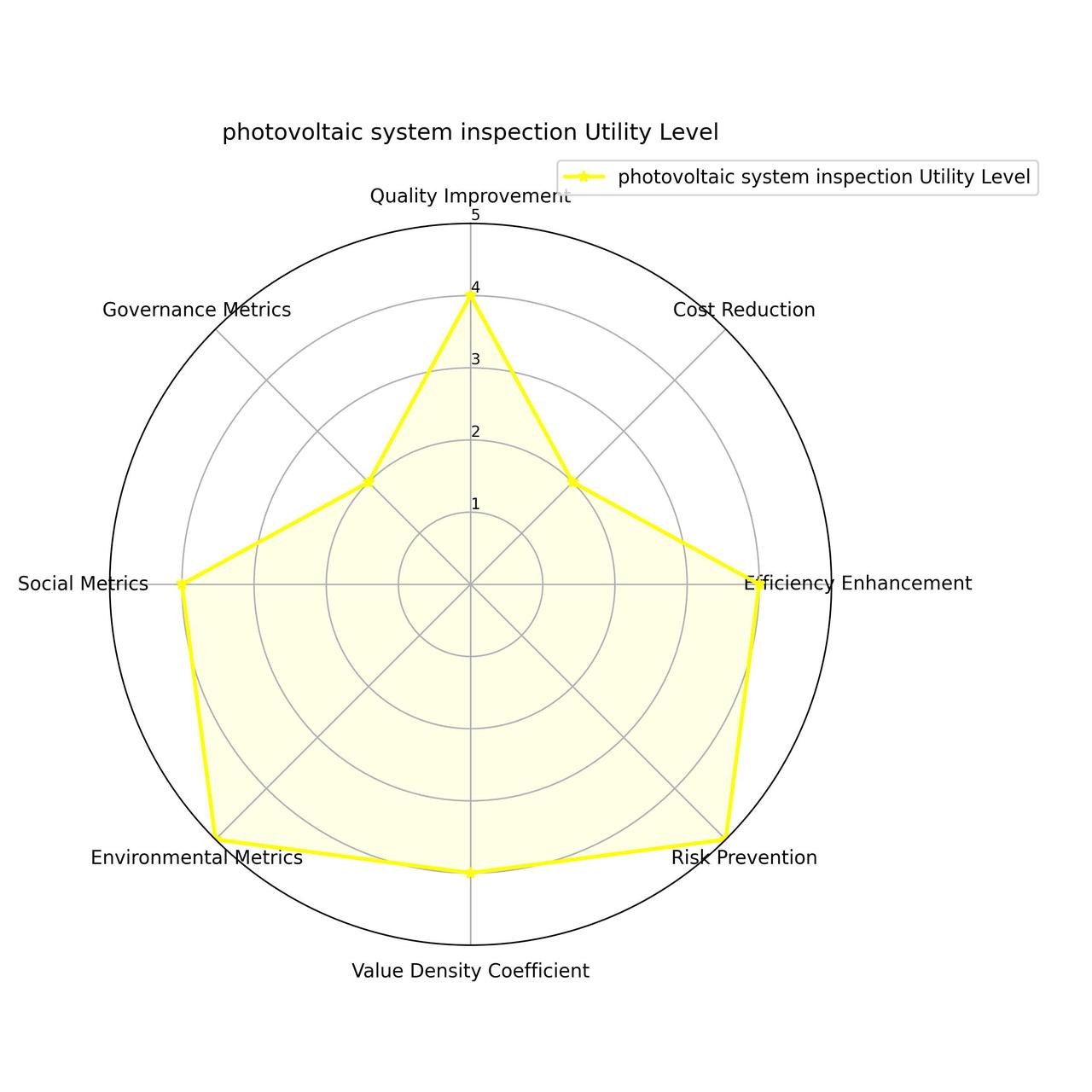}
        \caption{}
        \label{fig:pv_u_l}
    \end{subfigure}
    \caption{re-conversion Scores—Capability, Utility, and Adoption in PV inspection}
    \label{fig:pv_level}
\end{figure}

Figure \ref{fig:pv_level} presents expert panel ratings of multimodal AI in PV inspections across Capability, Adoption, and Utility dimensions, using a 1-5 scale. Figure \ref{fig:pv_c_l} shows capability assessment across 8 tertiary dimensions, Figure \ref{fig:pv_a_l} displays adoption potential across 9 dimensions during deployment, and Figure \ref{fig:pv_u_l} illustrates value realization across 8 dimensions in the PV inspection industry.

Expert panel assessment revealed strong capability ratings across multiple dimensions, with Environmental Adaptation receiving the highest possible rating (Level 5, score 100) due to the system's robust performance across varying weather conditions. Human-Machine Interaction Maturity likewise received exceptional ratings (Level 5, score 100), reflecting sophisticated interfaces designed for field technicians.
\subsubsection{TEMAI Evaluation Results for PV Implementation}
\setlength{\parindent}{0cm}
The TEMAI framework assessment yielded an initial Capability score of 70.19, reflecting the multimodal system's strong technical alignment with PV inspection requirements. The Adoption conversion rate was calculated at 65.23 \%, resulting in an effective Capability score of 45.78 after implementation constraints. The Utility value conversion rate reached 77.04 \%, producing a final value realization score of 23.01—more than double that of the retail implementation.

\begin{figure}[htbp]
    \centering
    \begin{subfigure}[b]{0.3\textwidth}
        \centering
        \includegraphics[width=\textwidth]{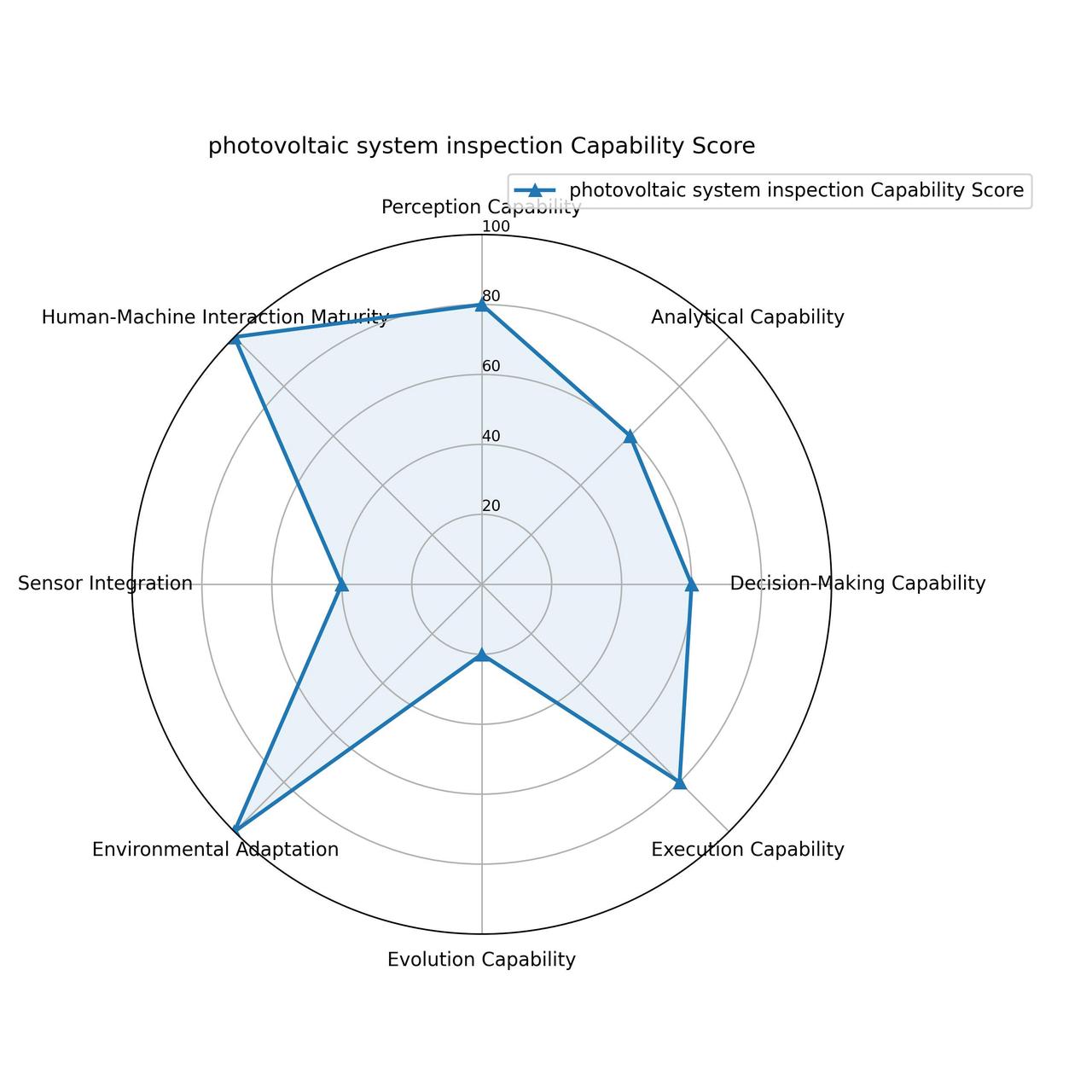}
        \caption{}
        \label{fig:pv_c_s}
    \end{subfigure}
    \hfill
    \begin{subfigure}[b]{0.3\textwidth}
        \centering
        \includegraphics[width=\textwidth]{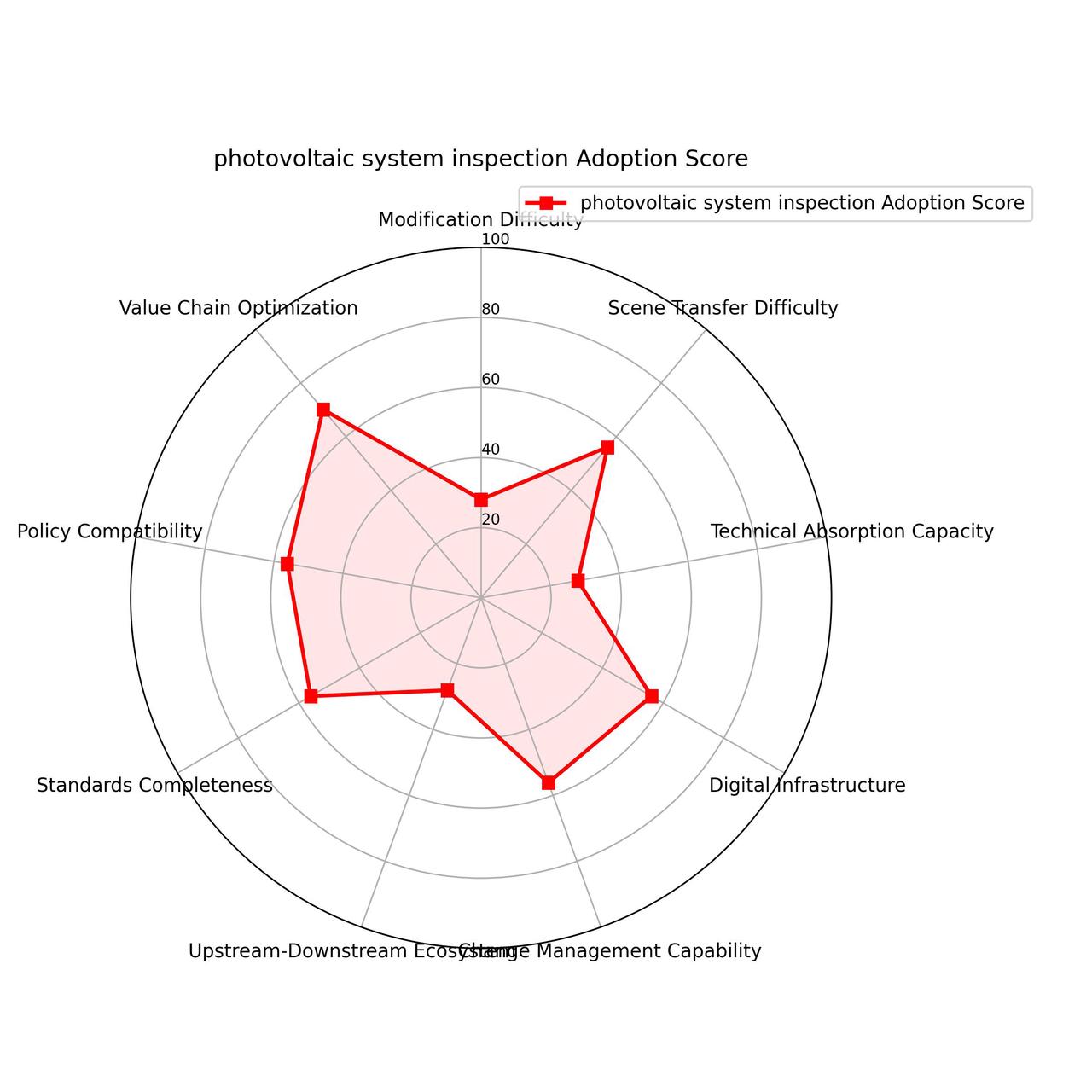}
        \caption{}
        \label{fig:pv_a_s}
    \end{subfigure}
    \hfill
    \begin{subfigure}[b]{0.3\textwidth}
        \centering
        \includegraphics[width=\textwidth]{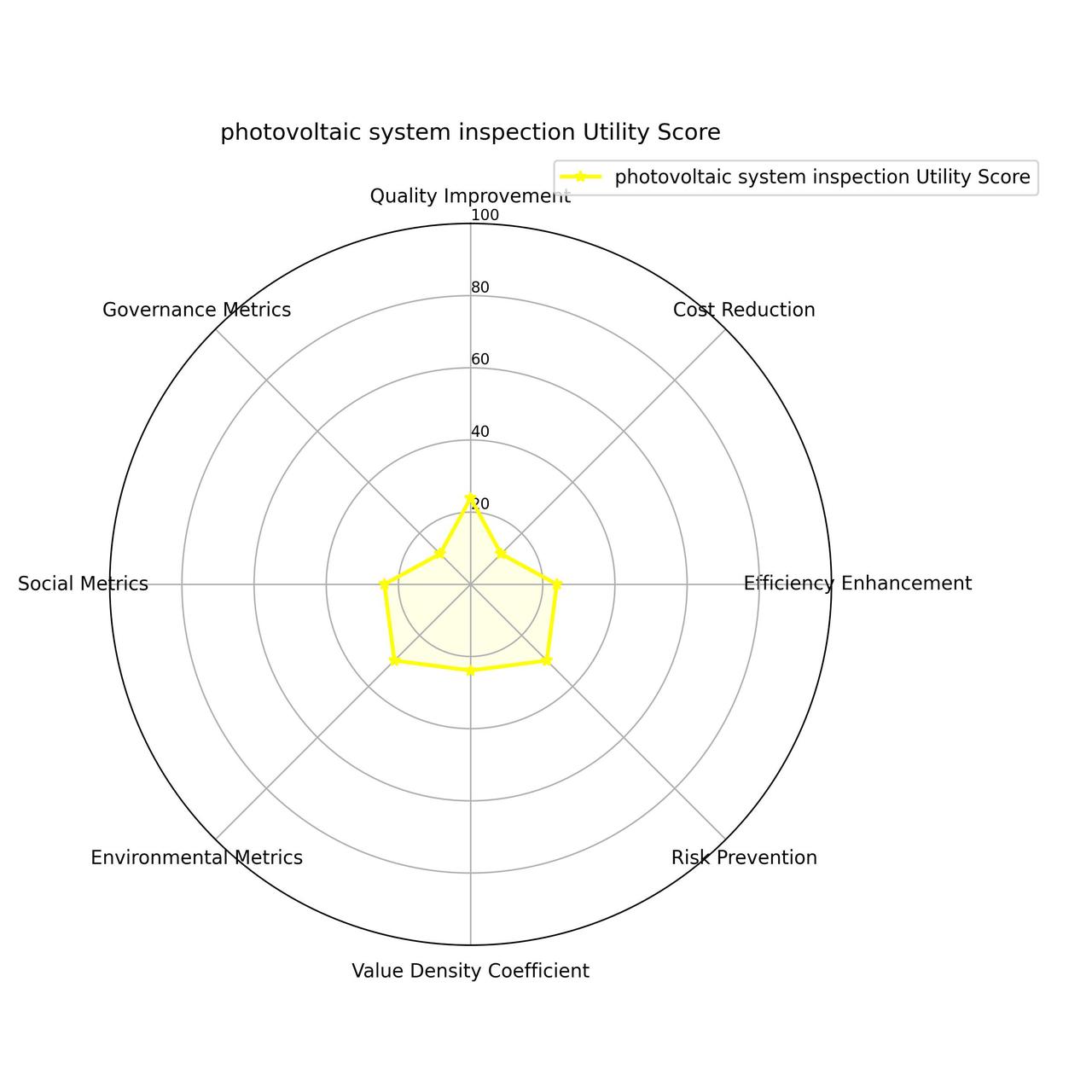}
        \caption{}
        \label{fig:pv_u_s}
    \end{subfigure}
    \caption{Converted Scores—Capability, Utility, and Adoption in PV inspection}
    \label{fig:pv_score}
\end{figure}

\setlength{\parindent}{2em}
Figure \ref{fig:pv_score} illustrates scores (0-100 scale) for multimodal AI in PV inspection across implementation stages. Figure \ref{fig:pv_c_s} shows the original AI capability scores, Figure \ref{fig:pv_a_s} displays transformed capabilities during deployment, and Figure \ref{fig:pv_u_s} presents the final value realization scores in the PV inspection industry.

Analysis of the component-level results revealed several notable implementation characteristics:

In the Capability dimension, Evolution Capability received the lowest score (Level 1, score 20), indicating limited self-improvement mechanisms in the initial deployment. Sensor Integration also scored relatively low (Level 2, score 40), highlighting integration challenges between thermal, visual, and electrical data streams in field conditions. However, Perception Capability (Level 4, score 80) and Execution Capability (Level 4, score 80) demonstrated strong performance in core inspection functions.

The Adoption dimension showed particular strength in Value Chain Optimization (Level 5, the converted score approximately 70), reflecting the system's effective integration into maintenance scheduling workflows. Policy Compatibility (Level 4,  the converted score approximately 65) and Standards Completeness (Level 4,  the converted score approximately  60) also scored well due to established regulatory frameworks for PV systems. However, Technical Absorption Capacity (Level 2,  the converted score approximately  30) and Modification Difficulty (Level 2,  the converted score approximately  30) indicated implementation barriers related to technical complexity and customization requirements.

The Utility dimension demonstrated exceptional performance in Risk Prevention (Level 5, the converted score approximately 30) and Environmental Metrics (Level 5,  the converted score approximately 30), highlighting the system's effectiveness in preventing catastrophic failures and supporting renewable energy objectives. Conversely, Cost Reduction (Level 2,the converted score approximately 10) scored lower, reflecting the high initial investment requirements relative to immediate operational savings.
\subsubsection{Implementation Enablers and Success Factors}
\setlength{\parindent}{0cm}
The PV case implementation revealed several critical success factors that enhanced implementation effectiveness:

\setlength{\parindent}{2em}
Established Safety Protocols: Pre-existing safety culture in the PV industry facilitated acceptance of AI inspection tools when framed as safety enhancements rather than replacements for human expertise.

Data Integration Architecture: Implementation success depended heavily on developing robust data pipelines connecting diverse sensor outputs with historical performance records to enable longitudinal analysis.

Risk-Based Prioritization: Implementing detection confidence scoring and consequence weighting algorithms allowed limited maintenance resources to be allocated optimally based on both detection certainty and potential impact.

Incremental Implementation Strategy: Starting with high-value application cases (hotspot detection) built organizational confidence before expanding to more complex inspection tasks (performance degradation analysis).

\subsection{ Comparative Analysis of TEMAI Implementation Across Domains}
\setlength{\parindent}{0cm}
The application of TEMAI across retail and PV inspection domains reveals significant differences in implementation pathways and value realization patterns, providing valuable insights into domain-specific adaptation requirements for multimodal AI systems.
\subsubsection{Capability Utilization Differences}

Comparative analysis reveals a 12.63-point difference in initial Capability scores (70.19 for PV versus 57.56 for retail), reflecting the different technical maturity of multimodal AI applications across these domains. This difference was most pronounced in Environmental Adaptation (100 versus 60) and Human-Machine Interaction Maturity (100 versus 60), indicating that PV inspection systems have developed more sophisticated environmental resilience and user interfaces.

\setlength{\parindent}{2em}
Notably, both domains showed similar Perception Capability scores (80), suggesting that core detection capabilities are reaching implementation maturity across diverse applications. However, Decision-Making Capability showed substantial divergence (60 for PV versus 40 for retail), indicating greater autonomy tolerance in PV settings where qualified human operators are scarce and inspection sites are often remote.

The capability utilization patterns reveal that retail implementations prioritize consistent basic functionality across diverse environments, while PV implementations focus on exceptional performance in specific high-consequence detection tasks.
\subsubsection{Adoption Barriers and Facilitators}
\setlength{\parindent}{0cm}
The Adoption dimension showed substantial variation between domains, with conversion rates of 65.23 \% for PV versus 51.16 \% for retail—a 14.07 percentage point difference that significantly impacted realized value. This discrepancy was primarily driven by differences in Scene Transfer Difficulty (L4 versus L2), demonstrating the greater standardization in PV installations compared to highly variable retail environments.

\setlength{\parindent}{2em}
Interestingly, Technical Absorption Capacity showed reversed performance (L2 for PV versus L4 for retail), indicating that despite the technical sophistication of the PV industry, retail organizations demonstrated superior capability to integrate AI inspection data into operational workflows. This counterintuitive finding suggests that technical complexity does not necessarily correlate with absorption capacity, which depends more on organizational data culture and existing analytical capabilities.

Value Chain Optimization showed the largest positive divergence for PV implementation (L5 versus L3), reflecting the system's effective integration into maintenance scheduling and lifecycle management processes.

\subsubsection{Value Creation Differentiation}
\setlength{\parindent}{0cm}
The Utility conversion rates (77.04 \% for PV versus 70.46 \% for retail) show that both implementations created substantial value, but with different patterns across components.

\setlength{\parindent}{2em}
Environmental Metrics demonstrated the most dramatic difference between applications (L5 for PV versus L2 for retail), highlighting the natural alignment between renewable energy infrastructure and environmental value creation—a pattern consistent with ESG investment priorities in these sectors.

The final value realization scores (23.01 for PV versus 10.61 for retail) indicate that the PV implementation achieved more than twice the value translation efficiency, demonstrating how industry-specific factors can dramatically influence the ultimate value derived from similar multimodal AI capabilities.

\subsection{Key Implementation Insights and Success Determinants}
\setlength{\parindent}{0cm}
The comparative application of TEMAI across different industrial inspection domains yields several critical insights that can guide future implementation strategies and maximize value realization.
\subsubsection{Implementation Strategy Alignment with Industry Characteristics}

Our cross-domain analysis reveals that implementation strategy must be fundamentally aligned with industry-specific characteristics rather than technical capabilities alone. The PV case demonstrated that industries with high standardization, critical safety implications, and clear regulatory frameworks provide more receptive environments for advanced multimodal AI systems, enabling higher adoption rates and value conversion efficiency.

\setlength{\parindent}{2em}
Conversely, environments with high variability and significant human factors, as seen in retail settings, require adaptive implementation strategies that emphasize human-AI collaboration rather than autonomous operation. These findings suggest that implementation planning should begin not with technical capabilities but with thorough domain analysis to identify alignment opportunities and potential barriers.

\subsubsection{Technical-Organizational Capability Balance}
\setlength{\parindent}{0cm}
Both implementations revealed that technical capability alone is insufficient for value realization. The PV case demonstrated higher technical capability scores but faced adoption challenges in Technical Absorption Capacity, while the retail case showed lower initial capability but stronger absorption characteristics. This pattern indicates that implementation success depends on balanced development of both technical capability and organizational readiness.

\setlength{\parindent}{2em}
The observed gap between initial capability scores and translated value (a reduction of 67.2 \% in retail and 67.7 \% in PV implementations) highlights the critical importance of organizational and adoption factors in determining ultimate value realization. This finding challenges the technology-first approach common in AI implementation and suggests that parallel investment in organizational capabilities may yield higher returns than pursuing marginal technical improvements.

\subsubsection{Implementation Pathway Framework}
\setlength{\parindent}{0cm}
Based on our comparative analysis, we propose a four-stage implementation pathway framework that maximizes value realization across diverse industrial inspection contexts:

\setlength{\parindent}{2em}
Value Density Mapping: Systematically evaluate application domains using the TEMAI framework to identify high-value-density opportunities where impact potential is most concentrated.

Capability-Adoption Alignment: For targeted applications, assess the gap between technical capability and organizational readiness, developing intervention strategies to address the most significant limiting factors.

Progressive Implementation: Begin with high-confidence, high-consequence applications that build organizational trust, gradually expanding scope as implementation experience accumulates.

Continuous Value Assessment: Implement ongoing value monitoring using the TEMAI utility metrics to guide iterative system refinement and identify emerging value creation opportunities.

This structured approach addresses both the technical and organizational dimensions of multimodal AI implementation, providing a comprehensive framework for translating theoretical capabilities into realized value across diverse industrial inspection contexts.
\section{Discussion}
\setlength{\parindent}{0cm}
The TEMAI framework represents a significant advancement in the evaluation and implementation of multimodal AI technologies in industrial inspection contexts. By applying translational research principles to bridge the gap between AI innovation and industrial application, this work addresses the critical need for structured knowledge transfer methodologies in high-stakes industrial environments.

\subsection{Translational Value and Implementation Insights}

Our adaptation of healthcare AI evaluation principles to industrial inspection demonstrates the versatility of translational research methodologies across domains. The multiplicative relationship between Capability, Adoption, and Utility dimensions creates a holistic evaluation mechanism that transcends traditional siloed assessment approaches, establishing TEMAI as a systematic knowledge transfer mechanism bridging research with practical implementation.

\setlength{\parindent}{2em}
The comparative analysis of retail and photovoltaic inspection implementations provides compelling evidence for the framework's adaptability across diverse industrial contexts. The retail case revealed how environmental variability significantly influences implementation success, while the PV case demonstrated how standardization enhances adoption rates. The substantial gap between initial capability scores and translated value (approximately 67 \% reduction in both implementations) provides empirical validation for the framework's emphasis on adoption factors as critical value determinants.

\subsection{Value Quantification and Domain-Specific Application}
\setlength{\parindent}{0cm}
The Value Density Coefficient provides a standardized mechanism for calculating the compound effects of AI implementations across varied industrial contexts, enabling organizations to justify AI investments with greater precision. The ESG Value Creation component represents a particularly significant advancement, capturing metrics that have traditionally been difficult to quantify.

\setlength{\parindent}{2em}
The dramatic difference in Environmental Metrics between PV inspection (Level 5) and retail implementation (Level 2) reveals how industry context fundamentally shapes value creation patterns. Similarly, the divergent patterns in Technical Absorption Capacity—with retail organizations demonstrating superior integration capabilities despite lower technical sophistication—challenges conventional assumptions about organizational readiness. These insights suggest that successful implementation depends on aligning AI systems with existing organizational capabilities rather than imposing standardized technical solutions.

\subsection{Regulatory and Support Context as Implementation Determinants}
\setlength{\parindent}{0cm}
Industries positioned in different regulatory-support quadrants face fundamentally different implementation challenges and opportunities, as shown in Figure 6. Sectors with high regulatory intensity combined with strong government support create optimal conditions for comprehensive AI inspection implementation, as compliance requirements align with available technological assistance. Conversely, highly regulated industries with minimal support encounter economic challenges that necessitate strategically focused implementations prioritizing critical inspection points.

\begin{figure}[htbp]
    \centering 
    \includegraphics[width=1\textwidth]{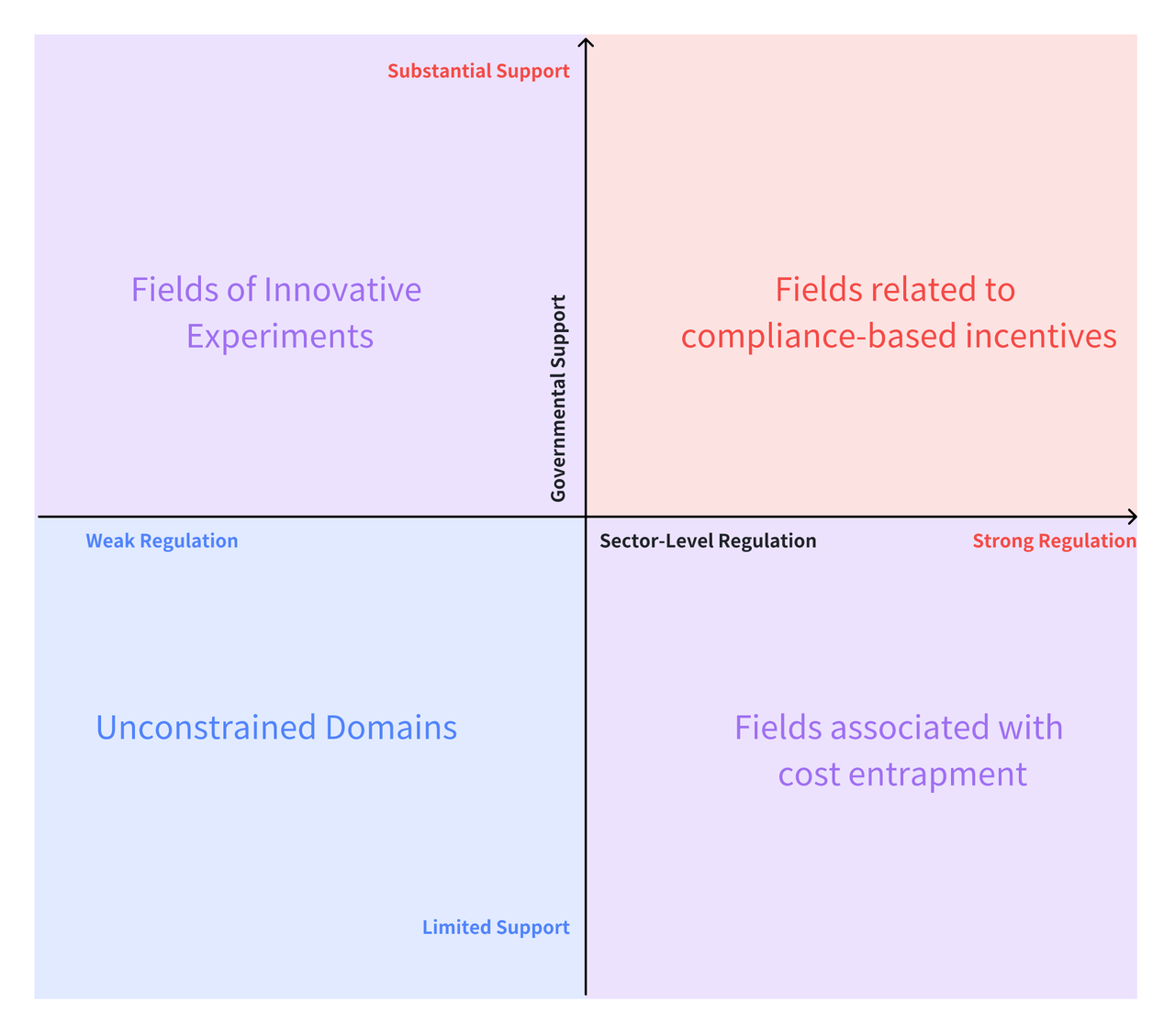} 
    \caption{Regulatory-Support Quadrant Analysis for TEMAI Implementation} 
    \label{fig:Regulatory Support Quadrant Analysis} 
\end{figure}

\setlength{\parindent}{2em}
The TEMAI framework's Adoption dimension, particularly through assessment criteria like Policy Compatibility and Standards Completeness, enables organizations to evaluate external readiness alongside internal capabilities. The case studies demonstrate how these factors manifest differently across industries—with the PV sector benefiting from established safety protocols that enhanced adoption despite technical complexity, while the retail environment faced implementation barriers despite technically simpler applications.

The four-quadrant regulatory-support analysis complements the TEMAI framework by providing strategic guidance for implementation prioritization and resource allocation. Organizations must align their technical approaches with their specific regulatory landscape, leveraging available support mechanisms while addressing compliance requirements—a critical success factor that transcends purely technical or organizational considerations.

\subsection{Implementation Challenges and Future Directions}
\setlength{\parindent}{0cm}
The case studies revealed implementation challenges beyond the original framework. In retail environments, store-to-store variability created adaptation requirements that diminished efficiency gains, while the PV implementation encountered integration difficulties with multi-sensor data streams. These practical challenges highlight the need for implementation strategies that anticipate domain-specific obstacles beyond theoretical framework dimensions.

\setlength{\parindent}{2em}
Despite its comprehensive structure, the TEMAI framework has limitations. Current value metrics may not fully capture long-term ecosystem benefits, and the framework may inadequately address regional variations in technology adoption patterns. The rapidly evolving regulatory landscape presents ongoing challenges for framework stability, particularly regarding data governance requirements. The implementation pathways derived from our case studies provide structured guidance but require further validation across additional industrial contexts.

Future research should explore more sophisticated temporal modeling for long-term value assessment, incorporate dynamic Policy Compatibility mechanisms, and expand multimodal assessment criteria for emerging sensory technologies. The observed balance between technical capability and organizational readiness suggests that future implementations should prioritize parallel development of both dimensions rather than pursuing sequential technical-then-organizational approaches—a significant departure from traditional technology-first deployment models that emphasizes the translational nature of successful multimodal AI implementation.

\section{Conclusion}
\setlength{\parindent}{0cm}
This research introduces the TEMAI framework, making several significant contributions to industrial AI implementation. First, we have successfully adapted translational research methodologies from healthcare to industrial inspection contexts, establishing a systematic approach for bridging laboratory AI innovations and AI applications. Second, our three-dimensional evaluation system integrates Capability, Adoption, and Utility dimensions into a comprehensive assessment that captures the multiplicative relationships between technical performance, organizational integration, and value creation. Third, we have developed specialized metrics including the Value Density Coefficient and Technical Absorption Capacity index, providing quantifiable measures for previously intangible aspects of AI implementation. Finally, our structured Implementation Pathways with Development, Deployment, and Discernment checks offer practical guidance for multimodal AI adoption.

\setlength{\parindent}{2em}
The empirical verification in retail and PV environments demonstrates the framework's effectiveness while revealing specific enhancement opportunities across all dimensions. The framework's modular design supports adaptation to diverse industrial sectors where multimodal inspection offers significant advantages. By providing a systematic methodology for evaluating and implementing multimodal AI inspection technologies, TEMAI addresses a critical gap in industrial digital transformation strategies and enables more effective knowledge transfer between AI research and industrial application.
\section{Appendix A: Calculation Formula for Converted Score}
\setlength{\parindent}{0cm}
$ScoreCapability = \Sigma capability_i \times weightcapability_i$

\setlength{\parindent}{2em}
$ScoreAdoption =  ScoreCapability \times \Sigma adoption_i \times weightadoption_i$

$ScoreUtility =  ScoreAdoption \times \Sigma utility_i \times weightutility_i$

\begin{table}[htbp]
    \centering
    \tiny
    \begin{tabular}{|>{\centering\arraybackslash}p{3cm}|>{\centering\arraybackslash}p{3cm}|>{\centering\arraybackslash}p{5cm}|>{\centering\arraybackslash}p{3cm}|}
        \toprule
        Core Dimensions & Components & Assessment Criteria & weight\_Assess(‱) \\
        \midrule
        \multirow{8}{*}{Capability} 
        & \multirow{5}{*}{Intelligent Level 4x1} 
        & Perception & 1888.89 \\ \cline{3-4}
        & & Analysis & 1666.67 \\ \cline{3-4}
        & & Decision & 1444.44 \\ \cline{3-4}
        & & Action & 1111.11 \\ \cline{3-4}
        & & evolvability & 555.56 \\ \cline{2-4}
        & \multirow{3}{*}{Equipment Compatibility} 
        & environmental adaptability & 1333.33 \\ \cline{3-4}
        & & sensor fusion degree & 1111.11 \\ \cline{3-4}
        & & HMI Maturity & 888.89 \\ \midrule
        \multirow{8}{*}{Adoption} 
        & \multirow{2}{*}{Task Adaptability} 
        & Process Reengineering Complexity & 1794.44 \\ \cline{3-4}
        & & Cross - Scenario Scalability & 1372.22 \\ \cline{2-4}
        & \multirow{3}{*}{Organizational Preparedness} 
        & AI literacy & 1025.00 \\ \cline{3-4}
        & & Digital Infrastructure & 1081.94 \\ \cline{3-4}
        & & Change Mgt Capability & 1025.00 \\ \cline{2-4}
        & \multirow{3}{*}{Ecosystem Maturity} 
        & Upstream and Downstream Ecosystems & 1041.67 \\ \cline{3-4}
        & & Standard Completeness Degree & 916.67 \\ \cline{3-4}
        & & Policy Fit & 541.67 \\ \cline{2-4}
        & Value Optimization Pathway & Value Optimization Pathway & 351.39 \\ \midrule
        \multirow{7}{*}{Utility} 
        & \multirow{5}{*}{Economic Value} 
        & Quality Enhancement & 1528.73 \\ \cline{3-4}
        & & Cost Displacement & 1999.11 \\ \cline{3-4}
        & & Efficiency Amplification & 1763.92 \\ \cline{3-4}
        & & Risk Prevention & 1175.94 \\ \cline{3-4}
        & & Value Density & 587.97 \\ \cline{2-4}
        & \multirow{3}{*}{ESG Valuation} 
        & Environmental Footprint & 1472.17 \\ \cline{3-4}
        & & Social Impact Quadrant & 883.30 \\ \cline{3-4}
        & & Governance Alignment & 588.87 \\ \bottomrule
    \end{tabular}
    \caption{weight of store}
    \label{tab:weight of store}
\end{table}

\begin{table}[htbp]
    \centering
    \tiny
    \begin{tabular}{|>{\centering\arraybackslash}p{3cm}|>{\centering\arraybackslash}p{3cm}|>{\centering\arraybackslash}p{4cm}|>{\centering\arraybackslash}p{3cm}|}
        \hline
        Core Dimensions & Components & Assessment Criteria & weight\_Assess(\%) \\
        \hline
        \multirow{8}{*}{Capability} 
        & \multirow{5}{*}{Intelligent Level} 
        & Perception Capability & 1844.97 \\ \cline{3-4}
        & & Analytical Capability & 1656.54 \\ \cline{3-4}
        & & Decision - Making Capability & 1619.26 \\ \cline{3-4}
        & & Execution Capability & 1167.17 \\ \cline{3-4}
        & & Evolution Capability & 490.05 \\ \cline{2-4}
        & \multirow{3}{*}{Equipment Compatibility} 
        & Environmental Adaptation & 1217.27 \\ \cline{3-4}
        & & Sensor Integration & 1127.70 \\ \cline{3-4}
        & & Human - Machine Interaction Maturity & 877.03 \\ \hline
        \multirow{8}{*}{Adoption} 
        & \multirow{2}{*}{Task Adaptability} 
        & Modification Difficulty & 1540.86 \\ \cline{3-4}
        & & Scene Transfer Difficulty & 1178.14 \\ \cline{2-4}
        & \multirow{3}{*}{Organizational Preparedness} 
        & Technical Absorption Capacity & 1265.94 \\ \cline{3-4}
        & & Digital Infrastructure & 1107.52 \\ \cline{3-4}
        & & Change Management Capability & 1186.55 \\ \cline{2-4}
        & \multirow{3}{*}{Ecosystem Maturity} 
        & Upstream - Downstream Ecosystem & 1226.74 \\ \cline{3-4}
        & & Standards Completeness & 1121.63 \\ \cline{3-4}
        & & Policy Compatibility & 689.63 \\ \cline{2-4}
        & Value Optimization Pathway & Value Chain Optimization & 683.00 \\ \hline
        \multirow{7}{*}{Utility} 
        & \multirow{5}{*}{Economic Value Creation} 
        & Quality Improvement & 1323.07 \\ \cline{3-4}
        & & Cost Reduction & 1823.28 \\ \cline{3-4}
        & & Efficiency Enhancement & 1613.84 \\ \cline{3-4}
        & & Risk Prevention & 1452.53 \\ \cline{3-4}
        & & Value Density Coefficient & 565.29 \\ \cline{2-4}
        & \multirow{3}{*}{ESG Value Creation} 
        & Environmental Metrics & 1718.29 \\ \cline{3-4}
        & & Social Metrics & 1002.36 \\ \cline{3-4}
        & & Governance Metrics & 501.34 \\ \hline
    \end{tabular}
    \caption{weight of PV}
    \label{tab:weight of PV}
\end{table}

\section{GenAI Applications Statement}
\setlength{\parindent}{0cm}
The following generative AI tools were utilized in this project: DeepSeek-R1, Grok3, Claude 3.7, Sonnet Thinking, Meta, and Monica. These tools were employed for tasks including writing, editing, brainstorming research ideas, and providing formatting assistance.

\end{document}